RESEARCH ARTICLE

# The *mre11* A470 alleles influence the hereditability and the segregation of telosomes in Saccharomyces cerevisiae

In-Joon Baek, Daniel S. Moss, Arthur J. Lustig*

Department of Biochemistry and Molecular Biology, Tulane University Medical Center, New Orleans, Louisiana, United States of America

* alustig@tulane.edu

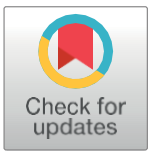

**Data Availability Statement:** Raw data for a real-time qPCR data is present at http://arxiv.org/abs/1705.06595.

**Funding:** This project was funded by NIH (www.nih.gov) R01GM069943 to AJL and by funds from the School of Medicine of Tulane University, Tulane University Dean of Research Office, and the Louisiana Consortium for Cancer Research. The funders had no role in study design, data collection and analysis, decision to publish, or preparation of manuscript.

## Abstract

Telomeres, the nucleoprotein complexes at the termini of linear chromosomes, are essential for the processes of end replication, end protection, and chromatin segregation. The Mre11 complex is involved in multiple cellular roles in DNA repair and structure in the regulation and function of telomere size homeostasis. In this study, we characterize yeast telomere chromatin structure, phenotypic heritability, and chromatin segregation in both wild-type [*MRE11*] and *A470* motif alleles. *MRE11* strains confer a telomere size of 300 base pairs of G+T irregular simple sequence repeats. This DNA and a portion of subtelomeric DNA is embedded in a telosome: a MNase-resistant non-nucleosomal particle. Chromatin immuno-precipitation shows a three to four-fold lower occupancy of Mre11A470T proteins than wild-type proteins in telosomes. Telosomes containing the Mre11A470T protein confer a greater resistance to MNase digestion than wild-type telosomes. The integration of a wild-type *MRE11* allele into an ectopic locus in the genome of an *mre11A470T* mutant and the introduction of an *mre11A470T* allele at an ectopic site in a wild-type strain lead to unexpectedly differing results. In each case, the replicated sister chromatids inherit telosomes containing only the protein encoded by the genomic *mre11* locus, even in the presence of protein encoded by the opposing ectopic allele. We hypothesize that the telosome segregates by a conservative mechanism. These data support a mechanism for the linkage between sister chromatid replication and maintenance of either identical mutant or identical wild-type telosomes after replication of sister chromatids. These data suggest the presence of an active mechanism for chromatin segregation in yeast.

## Introduction

The DNA-RNP structure present at the end of all eukaryotic chromosomes, the telomere, is necessary for end synthesis and protection. Our insight has grown into the polymerases, telomerase regulators, and chromatin components that solve the end-replication problem and contribute to the telomere end-protection [1]. The major mechanism to compensate for replicative attrition is the RNP-reverse transcriptase, telomerase [2]. Telomerase catalyzes the addition of G+T sequences onto pre-existing 3' telomere ends. The activity of several stable complexes, including telomerase, Ku70/Ku80 [3], CST [4–6], MRX [Mre11/Rad50/Xrs2][7, 8], Hoogsteen base-paired G4 DNA and G4 DNA binding proteins, protect the telomere [9–11]. A regulated competition between positive and negative factors creates a telomere homeostasis [12, 13] that in most eukaryotes acts as a dynamic telomere cap.

In all organisms, both Mre11 and Rad50 are dimeric, while NBS1 [or yXrs2] is monomeric. Together they form the MRX [or MRN] complex [14]. In the yeast model system *Saccharomyces cerevisiae*, the dynamic anti-checkpoint process begins with the binding of Tel1 [yATM] and Mre11 to only short telomeres [15–17]. Phosphorylation of telomerase by Tel1 results in the positive regulation of telomerase. In addition, Tel1 represses Rif1 [18], a protein that binds to the major telomere binding protein Rap1, and [19, 20] the single-strand DNA binding protein Cdc13 is activated by Cdk1[21, 22] phosphorylation. Both of these activities lead to a telomere size homeostasis through modulation of telomerase activity. These activation steps are in competition with Rif1 and Rif2, negative regulators of telomerase, that bind to Rap1 [5]]. Rif1/Rap1 association increases the abundance of Rif1 in elongated telomeres. In contrast, Rif2 interferes with the activity of Tel1, leading to repression of telomere addition [23, 24].

Two types of subtelomeric elements are associated with the telomere tract, one or more 5.5 [short] or 6.7 [long] kb Y' repeats, and the X-class telomeres that are also present telomere distal to the Y' elements. The size of Y' class telomeres can be easily determined by cleavage with XhoI, 870 base pairs [bp] from the telomere proximal Y' terminus [25].

Telomeric chromatin structure [termed a telosome], unlike telomeric DNA, is partially protected against micrococcal nuclease [MNase] digestion. In wild-type cells the protected DNA fragment has a mean size of 400 bp [26]. In contrast, the telomeric tract is 300 bp in wild-type cells, suggesting that the telosome protects both subtelomeric and telomeric sequences. The duplex portion of DNA is slightly larger than the telomere tract and is bound, primarily, by Rap1 at specific sites once in every 18 bp of telomere tract [27]. The Rap1 crystal structure indicates that telomeric DNA threads through a "pore-like" structure, distinct from the nucleosomal wrapping of DNA [26, 28]. Additional proteins that bind to the telomere are also components of the telosome. In addition to the Rif1 and Rif2, Rap1 physically associates with Sir2/Sir4, and Sir3 heterochromatin [29–31]. We have used several characteristics of the telosome: chromatin structure, telomere size, and Mre11 association with telomeric chromatin to probe the linkage between heritability and segregation.

We have tested whether telomere chromatin [the telosome] segregates by random or conservative means. Conservative segregation is the maintenance of the same form of chromatin on both replicated sister chromatids [32, 33]. This results in the heritability of the phenotype associated with telomeric chromatin. We are particularly interested in the factors that are needed to maintain conservative segregation. The best-known case of conservative segregation is some classes of heterochromatin [34] that are continuously modified through rounds of DNA replication. Hence the accessibility of the chromatin to modification will not simply vanish after replication and can be "reseeded" during replication [35]. Reformation of histone modifications can explain, in part, conservative segregation. The mechanism, however, is still speculative and may differ in various types of heterochromatin, including telomeric regions. Similar modifications of both histones and telomere proteins are also found at human telomeres and may influence telomere length during development in human cells (e.g.[36, 37]). Non-histone modifications or histone/telomere proteins loops (e.g.[30, 38]) may also play a role in the yeast telosome segregation.

The structure of the first eukaryotic crystal of Mre11 [39] has facilitated research in the fields of telomere repair, recombination, and telomere size homeostasis. The highly conserved A470 motif consists of 13 consecutive amino acids [470–482] [40, 41] that, based on the crystal structure, are close to the Mre11/Rad50 interface [39].

In this study, we have used one allele in the A470 motif, mre*11A470T*, as a marker to separate heritability and telomere chromatin segregation [40]. Telosomes containing the Mre11 or Mre11A470T proteins display strikingly different MNase profiles. A cell population that produces only the mutant allele requires more time of MNase digestion than wild-type cells to cleave telomeres from chromatin. We show here that the association of Mre11A470T with the telosome is less than wild-type association. However, we cannot rule out an additional possibility that Mre11 or Mre11A470T also act in trans to influence telosome structure. We have compared telomere structure and hereditability phenotypes conferred by alleles located at the endogenous site [also called genomic site] with genes integrated at an ectopic site [[Fig 1]]. We have determined that only the product of the genomic locus is incorporated in the telosome and leads to heritability and conservative telosome segregation.

## Materials and methods

### Strains

All strains used in this study are isogenic to W303a [Mat a *MRE11 leu2-3,112, ura3-1, trp1-1, HIS3, can1-100 rad5-535*] or W303 *a* [Matα *MRE11 leu2-3,112 trp1-1 ura3-1 ade2-1 his3-11,15 can1-100 rad5-535*] strains. KMM4 [Matα *mre11A470T trp1-1 leu2-3,112 ura3-1 his-11,153 can1-100 rad5-535*] is a backcross between IJ-9c [*MAT* α *mre11A470T$^m$ trp1-1 ura3-1 1his3-11,15 rad-5-535*] and W303a [40]. All strains were purified by sub-cloning prior to use. To construct the *mre11Δ* null allele, we PCR-amplified DNA from an *mre11*: *KanR* cassette in the yeast strain BY4741 [Res. Gen. 95400.BY4741] and used this cassette to conduct a one–step replacement of the *MRE11* gene with the *mre11*: *Kan$^r$* cassette in W303 strains. Metabolic and telomere size phenotypes verified the identity of the strain. The *MRE11* gene integrates at a non-genomic [ectopic] locus and can complement a genomic null allele.

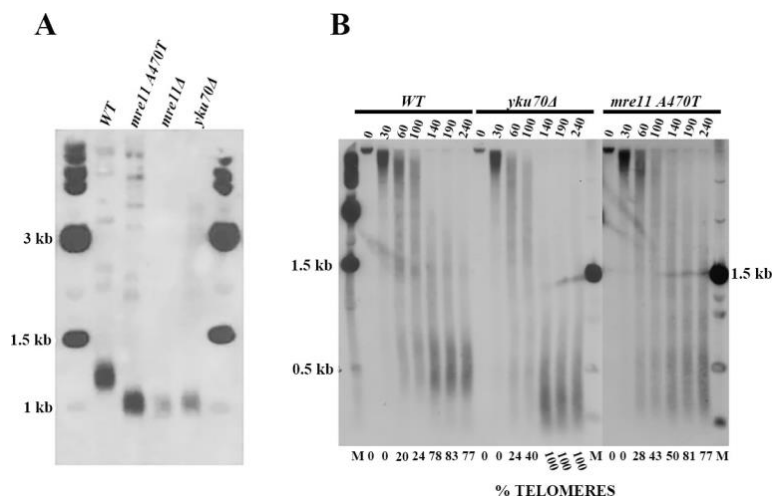

**Fig 1. Telomere and telosome phenotypes in *MRE11*, *mre11A470T*, *mre11Δ*, and *yKu70Δ* strains. [A]** Telomere size was determined by Southern blot analysis of *Xho*I-fractionated yeast DNA from *MRE11*, *mre11A470T*, *yku70Δ*, and *mre11Δ* cells using a telomere-specific probe. Wild-type strain had telomeres 300 bp in length, while both *mre11* mutants had telomeres centered at 150 bps. The *yku70Δ* telomeres were centered at 200 bp in length. The samples were flanked by λ size markers [NEB]. **[B]Telosome formation in *MRE11*, *yku70Δ*, and *mre11A470T* cells**. Chromatin isolated from wild-type, *yku70Δ*, and *mre11A470T* cells was digested with one unit/100 μl MNase for the times indicated, fractionated, and subject to Southern blot analysis using the A750 probe. The control was performed under the same conditions in the absence of enzyme. The position of 500 bp using the λ size marker is shown on the left and right of the gel. A gel splice is present due to the elimination of any empty lane adjacent to the *yku70Δ* set of data. The percentage of the signal that represents telomeres is shown below the gel.

https://doi.org/10.1371/journal.pone.0183549.g001

The *rap1-5* allele of *RAP1* was generated as described [42]. Previous studies indicated that the integrated *rap1-5* allele has a weak temperature sensitive [ts] phenotype at 37˚C [43]. To test the genetic interaction between *MRE11* and *RAP1*, we constructed the double mutant, *rap1-5 mre11A470T*. All growth conditions with *rap1-5* strains were conducted at 25˚C, the permissive temperature, or 30˚C, the semi-permissive temperature, respectively.

### Media

Strains in the absence of selection were grown in YPD media or YPD with adenine supplementation [YPAD], to a general facilitator of growth [44]. Strains containing the *Kan$^R$* cassette were grown in YPD + gentamycin. Strains auxotrophic for any metabolic marker was grown on synthetic minimal media containing the missing amino acid at standard concentrations [44].

### Rad5 function

We tested the possible involvement of the *rad5-535* allele, present in the W303 genetic background, in the generation of *mre11A470T* phenotypes. Ubiquitin is a major regulator of repair occurring during transcription [transcription-coupled repair]. Our concern derives from our use of another major repair complex, Mre11/Rad50/Xrs2, that is responsible the response to double-strand DNA damage repair and to some cases of residual transcriptional-coupled repair [45]. Whether there is a relationship between ubiquitin in transcription coupled repair and/or double strand break repair remains unknown. We therefore took a conservative approach and tested the *MRE11* and *mre11A470T* strains for their dependence on the *RAD5* gene. To test the possibility that *rad5-535* influences the *mre11A470T* allele, the telomere size of a *rad5: Kan$^R$ mre11A470T* double mutant was compared to single gene mutations. The short telomere length and heritability phenotypes of *mre11A470T* were not influenced by the *rad5-535* allele, indicating that the Rad5 is not involved in the telomeric Mre11A470T function. Wild-type strains containing only the *rad5-535* allele also had no influence on the telomere size phenotype and heritability [S1 Fig].

### Ectopic integration

The *EcoR1/XbaI* containing flanking DNA from *MRE11* and *mre11A470T* cells were cloned into pRS306 [46], creating pRS306-8 and pRS306-9, respectively. The *URA3* gene in pRS306 was cleaved by *NcoI* and transformed into strains carrying either *MRE11* or *mre11A407T* loci at the endogenous "genomic" locus [on chromosome XIII]. Integration resulted in Ura3$^+$ at the site of recombination at the ectopic site [also called an "ectopic allele" [e]] on chromosome V. In contrast, all endogenous genomic loci were Ura3$^-$ in the absence of the ectopic allele. All transformants contained a single copy of either *MRE11* or *mre11A470T*. Phenotypes were specified by the allele at the ectopic and genomic loci as e-*MRE11* g-*MRE11*[AJL551], and e-*mre11A470T* g-MRE11 [AJL552], e-*mre11A470T* g-*mre11A470T* [AJL553], and e-*MRE11* g-*MRE11* [AJL 554] strains.

### Ectopic assays for additional A470 motif alleles

The additional A470 motif alleles used in this study, *mre11V471A*, *mre11E479A*, *mre11K480A* and *mreK482T*, were placed at the endogenous site by two step gene replacement of W303 with the pRS306-cloned alleles. These strains were integrated by homologous recombination of the MRE11 gene into the ectopic site [ura3] to form e-*MRE11* g-*mre11A471T*, *e-MRE11* g-

*mre11E479A*, *e-MRE11*g-*mre11K480A*, *and e-MRE11*g-*mreK482T* [designated AJL555-AJL558, respectively].

After integration of each A470 motif allele, we routinely retested for the presence of the mutation. To that end, we used the Mre11 forward primer, [5'-CAAACGTATAGATAGATA TACCCAAT-3'], and the Mre11 reverse primer, [5'-GCTCCTCTCAAAATGGCATACCTT G-3'] for PCR amplification and DNA sequencing.

### Measurement of growth rates and viability

We measured the growth rates of independent *MRE11*, *mre11A470T*, *rap1*-5, and *mre11A470T rap1*-5 strains at three temperatures [23˚C, 30˚C, and 37˚C] in three trials. Cells were grown overnight at 30˚C and inoculated into 5 ml of YPAD media at the differing temperatures. The Cellometer Vision fluorescent counting system [Nexelcom Inc.] was used to visualize both viable and inviable cells. Propidium iodide is capable of entering and staining only inviable cells. Thus, viability is given as [[total number of cells–the number of propidium iodide-stained cells] /total cells] x100. Cell growth was measured in the absence of propidium iodide.

### Telosome analysis

We analyzed telosome structure as described [26, 47]. Specifically, we aliquoted nuclei into 100 microliters [µl] MNase buffer with 1 mM $CaCl_2$. A predetermined amount of MNase, based on preliminary studies, was incubated with each strain tested at 30˚C as a function of time [0, 30, 60, 100, 140, 190, and 240 seconds [secs], and the reaction was terminated. Negative controls were treated identically, but lacked MNase. The chromatin fraction of each culture was isolated and analyzed by Southern blot analysis using the 750 bp fragment, purified from plasmid Ap135 [A750, kind gift of Dr. Alison Bartuch], as the telomere probe. One of three size markers [the lambda size ladder, 1 [kilo-base] kb ladder and the 100 bp ladder [NEB] flanked each of the gels. Since chromatin experiments display subtle variations among independent experiments, we interpret the results of each experiment with all samples conducted in parallel on a quantitative basis, rather than the mean of data generated in differing experiments. Nonetheless, the pattern of size distributions is qualitatively reproducible between experiments.

### Telosome quantification

Each lane of one set of gels was scanned by the CLIQ software [Total Lab] and the fragments quantified using the "rubber band" background setting per manufacturer's instructions. High and low signals [in pixels] should be proportional to their abundance, since tract sizes are not the major variable in determining the increases in hybridization intensity [unlike measurements of telomere tract size]. The total pixels remain relatively constant [+/- 15%]. The intensity of telomere pixels/ the sum of all fragments gave rise to the % of the signal within the telomere size range. In some cases, the distribution varied slightly, and a broader range of a specific fragment was converted after increased treatment to a single species.

### Telomeric DNA isolated from telosomes

We isolated DNA from the chromatin fraction by phenol: chloroform extraction and ethanol precipitation. Samples were digested with *Xho*I to determine telomere size after Southern blot analysis using the A750 as a probe. We determined the telomere size using the Y' end-point at 870 bp from the *Xho*I site to terminus of Y' [48].

## Cycloheximide [CHX] protein stability assays

The CHX protein stability assay was performed in *MRE11* and *mre11A470T* strains. In the first day, the cells were pre-inoculated overnight in YPD and re-inoculated in YPD to newly prepared YPD media after overnight growth. The cells were grown to mid-log phase, and then 100 µg of CHX was added, incubated with CHX for 0, 15, 30, 60, 90, 120, 160, 150, 180, and 240 minutes [min] at 30°C. The proteins were extracted by the TCA method [49]. For Mre11 protein detection, a diluted [1/1000] primary Mre11 antibody [Rabbit polyclonal to Mre11-ChIP Grade, ABCAM ab:12159] was bound for one hour at RT and membrane was washed with TTBS buffer [a mixture of Tris-buffered saline [TBS] and Polysorbate 20 [Tween 20]] three times for 5 min at RT. The 1/2000 diluted secondary antibody [anti-rabbit, Santa Cruz sc:2313] was bound for one hour at RT and washed with TTBS buffer four times for 5 min at RT. The Mre11 proteins were detected by Western Blotting Luminal Reagent [Santa Cruz; sc:2048].

## Reverse transcriptase Real-Time PCR for Mre11 transcript abundance

RNA was extracted from *g-MRE11*, *g-MRE11e-MRE11*, and *e-mre11A470T g-MRE11* strains using phenol: chloroform: isoamyl alcohol and subsequent ethanol precipitation. The cDNA was synthesized from one microgram of RNA after addition of four µl of I-Script RT Supermix [Bio-Rad], and the mix was diluted to 20 µl with dH2O. After primer extension, the reaction was incubated in the Bio-Rad complete mix for five min at 25°C, and with reverse transcriptase for 30 min at 42°C. Reverse transcriptase was inactivated for one minute at 95°C. The qPCR, iTaq™ Universal SYBR **1** Green Supermix [Bio-Rad] was used for CFX96 Touch™ Real-Time PCR Detection System [Bio-Rad]. The primers used are, *MRE11* F: `ATGGTTGCGGAATTACCGA`, MRE11R: `CCAACTTCTGGTAATAAAGATAG`, ACT1 F: `GTCCCAATTGCTCGAGAGATTTCT`, and ACT1 R: `GACCATGATACCTTGGTGTCTTGG`.

## Chromatin immunoprecipitation [ChIP] methodology

We followed the procedure of Hecht et al [31] using the following modifications: Cultures were scaled to 20 ml, and cells were incubated with 1% formaldehyde for 15 min at 37°C before quenching. Cells were washed twice with cold PBS and suspended in collection of the lysate. Cells were sonicated three times for four secs at a setting of five [Dismembrator Model 100, Fisher]. The pellets were then centrifuged, and the supernatant containing the extract was collected. The immunoprecipitation used the ChIP-grade anti-Mre11 antibody [1/150] incubated overnight at 4°C on a rotator. We pre-equilibrated the protein [A/G] agarose beads [Santa-Cruz; sc2003] for two hours [hrs] at 4°C. We subsequently washed the beads twice with 500µl lysis buffer at room temperature$^m$. The fragments were eluted from beads with 50 µl 100 mM NaHCO$_3$, 1% SDS at 65°C for 15 min. Centrifugation separated the beads from the eluate. 50 µl TE and 100 micrograms [µg] proteinase K were added, and precipitated by ethanol. The final DNA was dissolved in 30 µl TE and subjected to PCR using primers flanking a 200 bp sequence to amplify the X/telomeric border using the X$_{forward}$ primer 5′-`GGAGCAACTTGCGTGAATCGAAGA`-3′ and the Tel$_{reverse\ primer}$: `CTGTCGATGATGCTGCTAAACTG`. $^m$Lysis buffer = [HEPES/KOH [50 milli-molar [mM] pH7.5], 500 mM NaCl, 1 mM EDTA, 1% Triton X-100, DOC [0.1%], SDS [0.1%] and a protease inhibitor cocktail [Santa Cruz; SC29131] dissolved in 30 µl TE.

## Real-Time PCR ChIP analysis of Mre11 abundance in telomeric chromatin

Cells were grown to OD600 = 0.2 in 20 ml of fresh liquid media to mid-log phase. Real-time qPCR was conducted according to the manufacturer's instructions [Bio-Rad] as summarized below. We cross-linked proteins in vivo using a 1% formaldehyde for 15 min in a shaker at room temperature. The formaldehyde reaction was then quenched with 125 mM glycine. The cells were washed twice with cold PBS and suspended in 400ul lysis buffer. Cells were lysed with glass beads in a bead-beater four times for 20 secs each, transferred to a new tube and sonicated twice for 10 secs. The lysate was collected in a new tube and centrifuged for 10 min at 4˚C. We immunoprecipitated the supernatant overnight at 4˚C with a 1/150 antibody dilution on a rotator. We subsequently added 25μl of a 50% slurry of the pre-equilibrated protein A/G. Agarose beads were incubated for two hrs on a rotator at 4˚ C. The beads were pelleted and washed with twice with 500 μl lysis buffer for two times at room temperature, once with 500 μl deoxycholate buffer beads [10 mM Tris ph7.5 0.25 M LiCl and 0.5% sodium deoxycholate and 1 mM EDTA] at room temperature, and once with 500 μl, TE for five min. The fragments were eluted from beads at 65˚C in 50 μl elution buffer for 15 min. The supernatant was collected and transferred to a new tube. The cross-links were reversed after incubating at 65˚C in 50 μl elution buffer overnight.

The beads were removed from the supernatant, and 50 μl TE and 100 μg proteinase K were added and incubated for three hrs at 55˚C. One hundred μl TE and 200 μl PCIA were added, mixed well and centrifuged for five min at room temperature. The supernatant was ethanol precipitated for 30 min at -20˚C, and the precipitate isolated after spinning at 14,000 x g for 30 min at 4˚C. The pellet was dried and dissolved in 300 μl TE. Procedures are accessible to the public by the MIQE convention [50].

We conducted real-time qPCR [Bio-Rad] using the CFX96 Touch™ apparatus and the primers Xforward and Telreverse described above. The data were analyzed using the Bio-Rad CFX Manager™ Software and the ΔΔC value was derived. Cq values from the CFX Manager software were used to determine the ΔCq [Target Cq- reference Cq], ΔCq expression [$2^{-\Delta Cq}$], mean ΔCq expression [average replicates], ΔCq expression standard deviation], and the ΔΔCq [normalized value to *mre11Δ*]. The ChIP analysis was carried out as before, and reverse crosslinking performed in a final volume of two microliters of ChIP DNA using two different dilutions of protein in 96 well plates that were subjected to qPCR.

### Telomere sequencing

We sequenced telomeres by the C-tailing method using the 5' `GTTCTCAGAAATTCT TCATT CGTAG-3'` [C-tail primer] and the unique telomere-proximal X-class primer 5` `-GAGTTGG ATACGGG TAGTTGG-3'` that shares homology with 13 termini [51]. Sequencing was carried out by Fasteris, Inc. [https://www.fasteris.com].

## Results

### Heritability and telosome structure in *mre11A470T* cells

**The altered telosomes of *mre11A470T* alleles.** *MRE11* and *mre11A470T* telomere tract lengths were measured by gel electrophoresis of *Xho*I sensitive cleavage Y' elements that are 870 bp proximal to the telomere. The lengths were 300 bp and 150 bp, respectively, in Y' and X [40] classes and packaged in telomere chromatin particles, termed telosomes [Fig 1]. After MNase treatment of chromatin, wild-type cells protect a telosome DNA fragment of 400–600 bp, as reported [26]. In contrast, the *mre11A470T* allele telosome is more resistant to MNase and protects a more diffuse pattern of subtelomeric sequences [100–900 bp]. Quantifying the signals leads to the conclusion that the *mre11A470T* telosome is present at maximal abundance after 240 secs of digestion, rather than 140 secs for wild-type telomeres. These data indicate that the telosome extends into subtelomeric sequences.

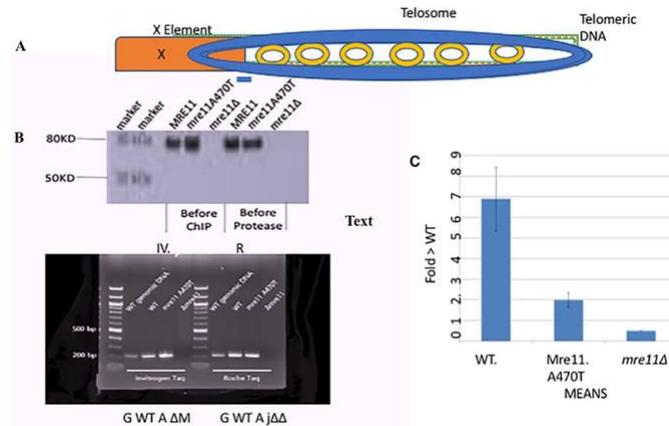

**Fig 2. Analysis of Mre11 and Mre11A470T association with telomeric chromatin of X-class telomeres.**
{**A**] Diagram of the telomere/subtelomeric region showing telomere sequence [patched], the telosome [blue], individual RAP1 molecules [yellow], subtelomeric sequence [orange]. Rif1 MRX and Tel1 **are also present**. The proteins binding to the single-stranded overhang are not shown. The 200 bp diagnostic species is shown in blue below the telosome and is generated by PCR. [**B**] Western analysis of cell extract isolated from strains carrying ectopic copies of *MRE11* or *mre11A470T* in an *mre11*: *Kan I$^R$* allele before and after ChIP [left and right lanes] using the anti-Mre11 antibody [Abcam]. [The 200 bp subtelomeric fragment adjacent to the telomere was amplified by an Invitrogen [IV] or Roche Taq polymerase after ChIP. PCR was performed with the Mre11 protein at saturating levels [as shown in Western]. The DNA sources are listed: G, genomic WT DNA; M, ChIP Mre11; A, ChIP *mre11A470T* and Δ, ChIP *mre11Δ*. [**C**] Summarized Real Time qPCR data. The qPCR technique is described in Materials and Methods. Two experiments were conducted, one in duplicate and one in triplicate [see Materials and Methods]. Abundance was calculated using the ΔΔC technique. As required by the MIQE convention [50]], raw data and calculations for these experiments are presented at the arXiv at http://arxiv.org/abs/1705.06595.

https://doi.org/10.1371/journal.pone.0183549.g002

**Micrococcal nuclease resistance phenotype.** The *mre11A470T* allele is more resistant than wild type to MNase. In the *mre11A470T* allele, extended times of digestion are required to reach the point of telosome release from chromatin. These data imply that the structure or folding of Mre11A470T and associated proteins protect telosomes from digestion. The delay in cleavage of *mre11A470T* chromatin is not a consequence of short telomeres, since *yku70Δ* telosomes has a wild-type level of resistance [**Fig 1B**]. In contrast, the *mre11* null allele confers a variety of chromatin structures from wild type to mutant, and may reflect multiple states of the telosome in the absence Mre11. As expected, DNA isolated from the chromatin fraction gave rise to the initial telomere sizes [**S2 Fig**].

**The effect of altered telosomes on telomere sequence.** A change in the relative position of the telomeres may be the consequence of Mre11A470T-induced changes in the TLC1 telomerase RNA template region. The telomere sequence could be altered or constrained in the mutant strain. We therefore sequenced multiple independent clones of both wild type and *mre11A470T* strains. However, we find no changes in sequence or sequence structure]. The sequences of both the wild-type, [**S3 Fig**] and *mre11A470T* DNA [**S4 Fig**], vary between subclones, but follow the yeast consensus sequence $G_{1-3}T$. Wild-type telomeres were conserved in the first 150 bp sequences distal to the terminus, but varied significantly in the terminal half due to deletions and sequence mismatches.

**ChIP analysis of *MRE11* and *mre11A470T* strains.** To determine whether the Mre11 and Mre11A470T proteins are capable of binding to the telosome, we conducted a simple chromatin immunoprecipitation [ChIP] analysis using excess Mre11 and Mre11A470T [41]] [**Fig 2A, top**]. Mre11 proteins were present as intact species of known size [80 kD] after Western analysis using the monoclonal antibodies directed against a C- terminal peptide of Mre11 [Abcam].

We crosslinked proteins in strains containing *MRE11* or *mre11A470T* alleles, and we immunoprecipitated cellular extracts with the same anti-Mre11 antibody. After elution, de-crosslinking, and protease treatment, a DNA was released that could support the PCR amplification of a 200 bp product at the junction between the X element and telomeric sequences. This fragment is maintained within the telosome, based on telomere size analysis. The *mre11Δ* allele, on the other hand, generated no protein species or amplified DNA after ChIP analysis [Fig 2B]. These binding data reinforce that both Mre11 and Mre11A470T proteins are a part of the telosome.

To estimate the relative binding of Mre11 and Mre11A470T to the telosome at X class telomeres, we performed real-time qPCR ChIPs [Fig 2C]. DNA isolated from chromatin that was immunoprecipitated by anti-Mre11 [ABCAM] in *MRE11*, *mre11A470T*, or *mre11Δ* extracts. The immunoprecipitated DNA was subject to real-time PCR to quantify the 200 bp junctional fragment [as a proxy for telomeres] using the same PCR primers as described above. We found a 3-4-fold increased abundance of telosomes containing wild-type Mre11 over Mre11A470T from wild-type and *mre11A470T* cells.

An alternative explanation for the ChIP data is a four-fold decrease in the stability of the mutant protein. To test this possibility, we conducted a cycloheximide [CHX] translational block, and examined the subsequent time points for protein degradation. Both wild-type and mutant Mre11 proteins had identical high stabilities, similar to the values of the actin loading control [S5A Fig]. We can conclude therefore that both proteins are bound and stable at the telomere chromatin.

## Phenotype and structural characteristics of alleles at ectopic and genomic positions

**Epistatic relationship between *MRE11* and *mre11A470T* alleles in heritability and chromatin structure.** In considering the ectopic/genomic system, we first had to resolve whether the allele at an ectopic site is silenced. Several lines of evidence argue against this possibility. First, a TAP-tag derivative of both wild type and *mre11A470T* was integrated at an ectopic site and produced proteins that are also stable in a CHX assay [S5B Fig]. Second, the allele at the ectopic site can be expressed, since the wild-type allele at the *URA3* locus on chromosome IX can complement an *mre11*: $Kan^R$ allele at the genomic site [see 4, right]. Third, based on real-time qPCR analysis, reverse transcription of genes located at both genomic and ectopic sites have a three- to four-fold increase over either strain carrying the genomic allele alone [S6 Fig]. These data suggest that the genes at ectopic and genomic sites are both transcriptionally active. Hence, the ectopic sites produce transcripts and intact tagged proteins.

Strains carrying the wild-type gene at the endogenous genomic locus [g] were transformed and integrated with the *mre11A470T* gene at an ectopic [e] *ura3* locus. Conversely, strains carrying the *mre11A470T* allele at the genomic locus were integrated with *MRE11* at the ectopic *ura3*. The only phenotypes observed were the result of the gene location at the endogenous locus [Fig 3].

Both the telomere tract length and MNase digestion pattern of chromatin were analyzed in e-MRE11 g-*MRE11* and e-*mre11A470T* g-MRE11 strains. In the opposing experiment, strains were analyzed in e-mre11A470T g-MRE11 and e-mre11A470T g-mre11A470T. We will term this type of strain as "ectopic/genomic" to reflect the positions of the two alleles. The heritability of short mutant telomeres in e-*MRE11* g-*mre11A470T* and g-*mre11A470T* strains persisted over 100 generations [the longest period tested]. In contrast, the wild-type telomere lengths persisted in e- *mre11A470T g-MRE11* and g-*MRE11* strains [Fig 4].

We found similar MNase patterns and spacing [diffusion of chromatin] isolated from e-*MRE11* g-*mre11A470T, and* e-*mre11A470T* g-*mre11A470T*, and isolated from e-*mre11A470T*

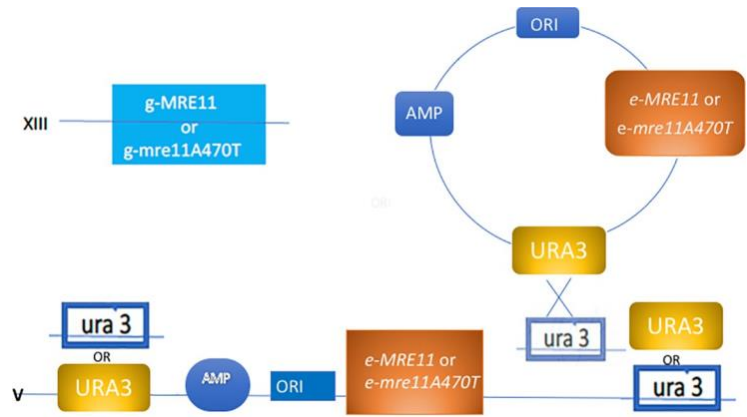

**Fig 3. Scheme for the formation of ectopic/genomic strains.** The ectopic copies of *MRE11* or *mre11A470T* were integrated at the *ura3* locus [clear] by homologous recombination using pIB306 integrating vector. This vector carries the *URA3* [yellow] at chromosome IX, either the *MRE11* or mre*11A470T* [orange] gene, and the beta-lactamase gene [the AMP gene] that confers resistance to [ampicillin] in E. coli. These plasmids were integrated into strains carrying the *mre11* or *MRE11* genomic allele [light blue], respectively. The DNA duplex is drawn as a single line. The Figure is not drawn to scale.

https://doi.org/10.1371/journal.pone.0183549.g003

g-*MRE11* and e-*MRE11* g-*MRE11* cells sub-cultured for the same number of generations [Fig 5]. This is similar to single mutants *mre11A470T* [maximal extent: 100 bp-900 bp] and *yku70Δ* [maximal extent 100 bp-900 bp] telosomes [Fig 1B]. In one case, an *mre11A470T* gene that integrated into an ectopic site in a *MRE11* strain [Fig 5, [e-mre11A470T gMRE11; 240 min]] produced a "diffuse" wild-type telosome [maximal extent: 200 bp-1000 bp]. Although insufficient to make a statistically sound conclusion, the diffusion phenotype is biased towards short telomeres. These diffuse telomeres are likely to represent a broader range of MNase cleavage sites that are less defined to a specified size range. Thus, in most aspects, the final phenotype is close to the genomic locus, regardless of the presence of opposing alleles at ectopic sites. As a control, *e-MRE11* complemented a null allele.

Assays of additional A470 alleles reveal their dominant epistasis when the A470T allele is at the genomic site and the *MRE11* allele is at the ectopic site. Each A470 motif allele confers either shorter telomeres [V471A, E479A], that [similar to the *mre11A470T* allele,] [Fig 6] or larger intermediate telomere sizes [K480A, A482T] that have equivalent telomere sizes in the presence or absence of the wild-type allele. These data demonstrate that a common trait of A470 alleles confine the phenotypic expression to the genomic loci.

### Rap1 and Mre11 interaction

**Rap1 and Mre11 mutant proteins confer synthetically lethal at 37°C.** We have shown that Mre11 can be defined both as part MNase nuclease sensitivity and by ChIP analysis. The telosome has previously been characterized as a component of the telomeric chromatin [26, 47, 52]. Although we do not know the relative degree of association of each component, we will assume they are generally part of the same telosome particle. We therefore tested the genetic or physical association of Rap1 and Mre11 at a basic level. To this end, we examined the growth and viability characteristics of a double mutant containing the *mre11A470T* and a weak temperature sensitive [ts] allele of *RAP1*, *rap1-5*. The *rap1-5* was chosen, in place of more severe *rap1* alleles, since *rap1-5* defects would allow changes in double mutants to be observed before lethality. *MRE11*, *RAP1*, *MRE11 rap1-5*, *mre11A470T RAP1*, and *mre11A470T rap1-5* cells were assayed for both cell growth viability and cell growth rate. Viability [as measured by

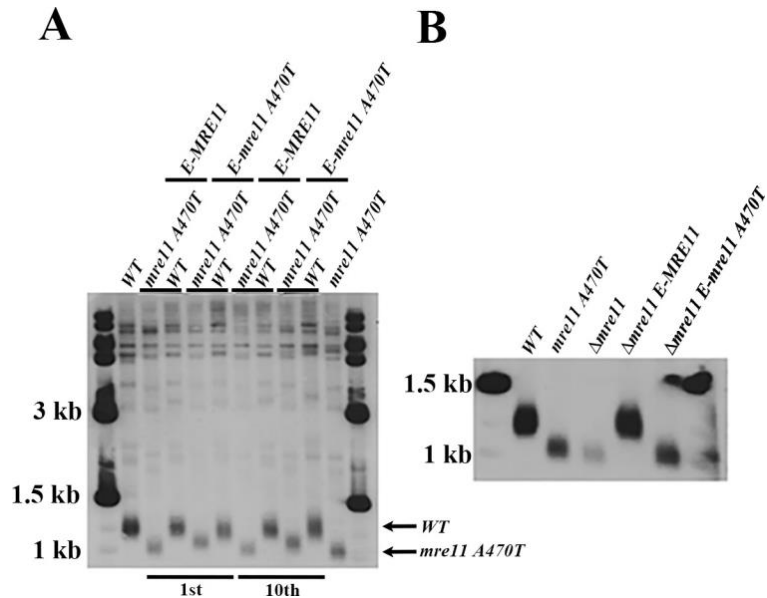

**Fig 4. Telomere size and heritability in ectopic/genomic strains. [A]** Telomere size homeostasis in e-*MRE11* g-*mre11A470T*, e-*mre11A470T* g-*MRE11*, e-*MRE11* g-*MRE11* and e-*mre11A470T* g-*mre11A470T* strains before [left] and after 100 PDs [right] [10 cycles of continuous subculturing with 10 population doublings each] of growth. *MRE11* [lane 2] and *mre11A470T* [lane 11] were also used as controls before and after sub-culturing. Markers flank the samples. Arrows designate the wild-type and mutant length of the telomeres. The identity of samples that were derived from DNA before and after sub-culturing are designated beneath the gel. **[B]** To check whether the wild-type gene at an ectopic site still behaves phenotypically as a wild type, despite an *mre11Δ* null allele, genomic DNA was extracted by phenol: chloroform: isoamyl alcohol extraction, precipitated with ethanol, digested with *Xho*I and subjected to Southern Analysis using the A750 probe. The e-*MRE11* allele complements the short telomere phenotype of a *mre11Δ* strain, indicating that the ectopic site is expressed.

https://doi.org/10.1371/journal.pone.0183549.g004

the % of viable cells] was tested for cells at 23˚C, 30˚C and 37˚C, reflecting the permissive, semi-permissive and restrictive temperatures of *rap1-5*. Viability was not altered at 23˚C and 30˚C in any strain, while *mre11A470T* grown at 37˚C decreased viability to 90%, and the double mutant to 60% [Fig 7]. [The *rap1-5* gene did not confer a viability phenotype over this

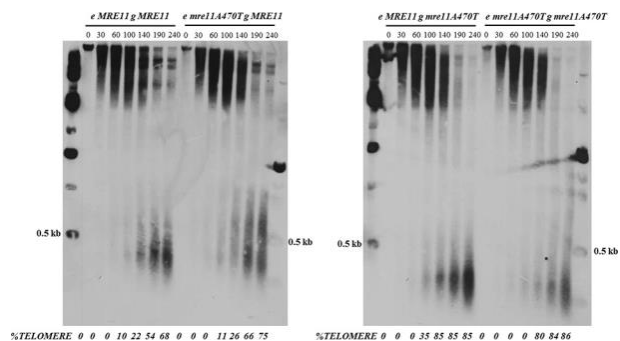

**Fig 5. Chromatin characterization from ectopic/genomic strains.** Telosome formation of e-*MRE11* g-*MRE11*, e-*mre11A470T*, g-*mre11A470T*, e-*MRE11* g-*mre11A470T*, e-*mre11A470T* g-*MRE11*, and the controls *MRE11* and *mre11A470T* were prepared as described in 2. The chromatin was extracted and digested by one unit /μl MNase at the designated time points. The gel was subjected to Southern blot analysis using A750 as the probe. The percentage of the signal that represents telomeres is shown below the gel lanes.

https://doi.org/10.1371/journal.pone.0183549.g005

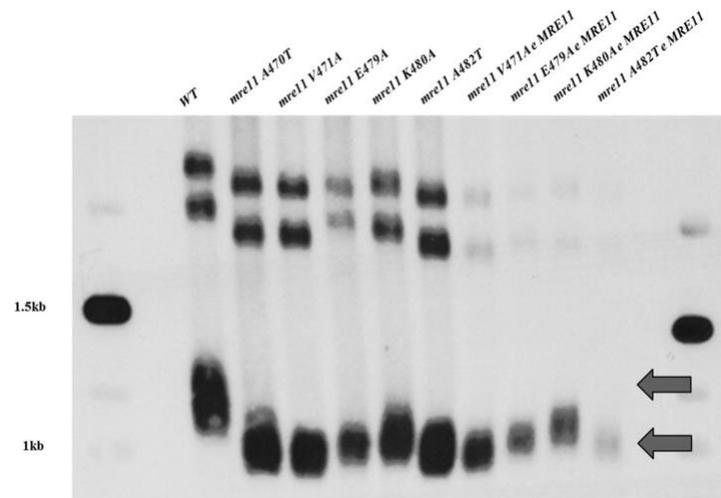

**Fig 6. The epistatic dominance of the A470 motif alleles at the genomic locus to wild-type.** Several additional A470 motif alleles, *mre11V471A*, *mre11E479A*, *mre11K480A* and *mre11K482T* in the presence of *MRE11* at the ectopic *ura3* locus. A Southern blot analysis of *Xho*I-digested genomic DNA, defining the Y' class of telomeres, was used to determine telomere size. A470 motif mutant strains could not be complemented by e-*MRE11*, as is the case for the mre*11A470T* allele. Wild-type strains were used as controls. Arrows point to wild-type and mutant strains that were used as controls.

https://doi.org/10.1371/journal.pone.0183549.g006

short time interval]. This non-epistatic relationship between the viability conferred by the phenotypes of *mre11A470T* and *rap1-5* suggests that Rap1 and Mre11 act independently on the telomere through different pathways or form a toxic complex that decreases viability.

Rap1 and Mre11 pathways influence viability [Fig 7] and growth rate [Fig 8] in distinct fashions. Growth rates were more significantly affected. At all temperatures, *mre11A470T rap1-5* and *mre11A470T* cell growth rates were reduced, with the double mutant having the most severe phenotype at each temperature [Fig 8]. The *rap1-5* strain showed no change in viability at any temperature within this time frame [Fig 7]. Only the *mre11A470T* and *mre11A470T rap1-5* strains were decreased in viability, albeit after differing times. Yet, cells stop growth before viability is lost, suggesting a window in which other events after growth cessation can still influence viability. Given the proteins involved, it is likely that either the lack of viability or the length of the cell cycle is the consequence of telomere-mediated genomic instability.

**Chromatin structure of *rap1-5 mre1A470T* telomeres.** To correlate growth rate with the telosome structure of the double mutant, we conducted a Southern blot of the products of MNase digestion in wild-type, *mre11A470T*, *rap1-5* and *mre11A470T rap1-5* cells grown at the semi-permissive temperature, 30°C [Fig 9]. Both the *mre11A470T rap1-5* and *mre11A470T* chromatin are relatively resistant to MNase until 100–240 secs, when both rapidly form telosomes. Wild-type and *rap1-5* cells start to form telosomes at an earlier time than *mre11A470T*. Hence, *rap1-5* does not facilitate more resistance to MNase. This is consistent with the small decrease in growth rate at 30°C in *mre11A470T* cells and the double mutants. We are somewhat disadvantaged by the inability to analyze chromatin structure at 37°C, given the high levels of unviability. This difference in sensitivity of different samples to reaction conditions of a single experiment is less than the difference in sensitivity to inherent variability between experiments. Nonetheless, the same qualitative pattern is maintained between identical experimental trials.

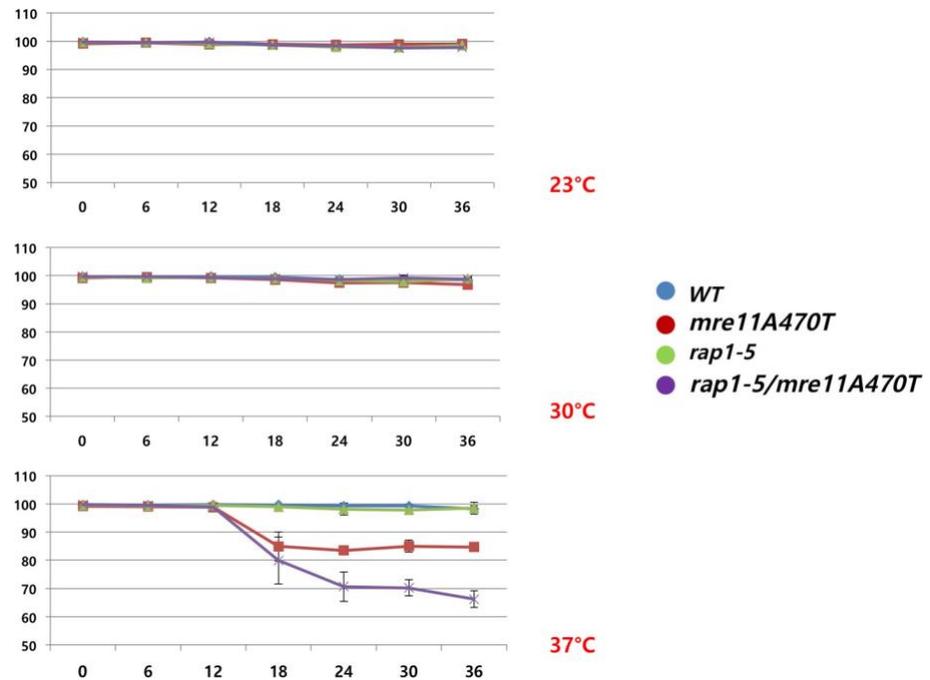

**Fig 7. Viability of *MRE11*, *mre11A470T*, *rap1-5*, and *mre11A470T rap1-5*.** 2.25x10$^6$ cells of a mid-log culture was centrifuged, placed into fresh media, and grown at 23°C, 30°C, or 37°C. Cell viability was assayed at the designated time points, and both Propidium iodide [inviable] stained cells and the total cell count were determined using the Vision Cellometer. *mre11A470T* and *mre11A470T rap1-5* cells had decreased viability only at 37°C. Viability decreased until 24 hours.

https://doi.org/10.1371/journal.pone.0183549.g007

## Discussion

Several lines of evidence indicate that telosome structure, hereditability and segregation are aberrant in the presence of the mre*11A470T* allele. First, an extended time of MNase cleavage is required to generate the Mre11A470T telosome. Real-time qPCR has shown a four-fold greater abundance of Mre11 than Mre11A470T in unsynchronized cells. This difference is not due to variations in protein stability.

Second, the mutant may not just alter structure, but could also facilitate the influence of a chaperone or regulatory RNA, such as *hsp90* [53]] and TERRA [[54, 55]], respectively. TERRA in yeast is currently an experimental 'black box' that may be the key to understand higher-level G-4DNA/telomeric DNA/ complexes [56].

Third, we have previously shown that the *mre11A470T* allele favors recombination over telomerase activity at chromosomal telomeres, based on the inability to heal a telomere seed sequence [41]. Indeed, *ctf18 mre11A470T* cells can confer a complete loss in telomerase activity *in vivo* [41]. This finding is consistent with the bypass of telomerase senescence in *mre11A470T tlc1Δ*, the apparent consequence of elevated telomere recombination [35]. One speculative possibility is that *mre11A470T* requires a cofactor to act directly on recombination. Alternatively, the structures of 3' termini may be a deciding factor in the activation of telomerase through the binding of different factors that influence heritability, telosome segregation, and genomic stability.

Finally, since Mre11 is dimeric, the conservative segregation pattern of telosome observed in *mre11A470T* cells must be related to the Mre11 and Mre11A470T homodimer and heterodimer stability. Alternatively, the absence of a chromatin modification may prevent the

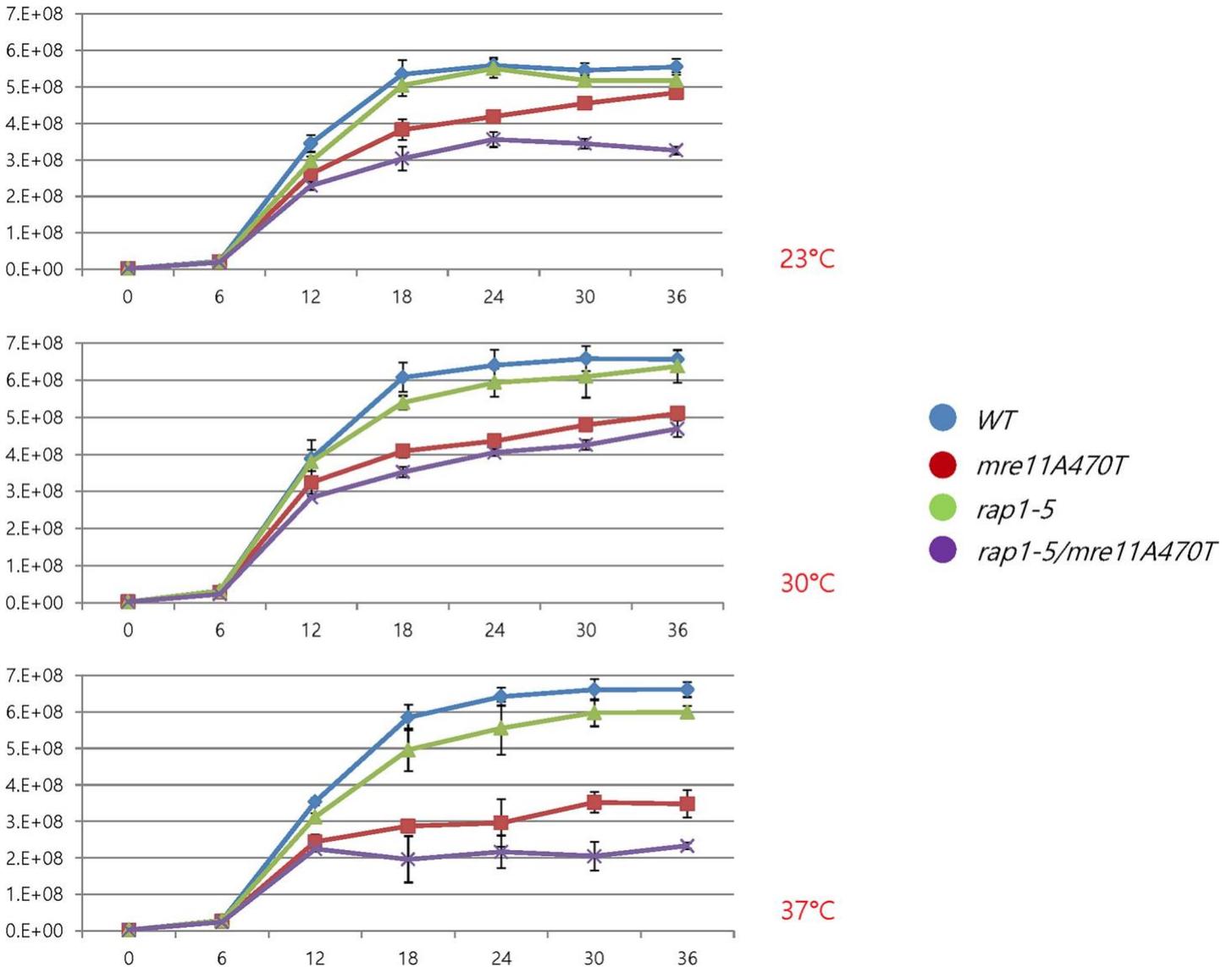

**Fig 8. Growth rates of *MRE11*, *mre11A470T*, *rap1-5*, and *mre11A470T rap1-5*.** The growth rates of the identical culture as in Fig 7. Growth rates are shown at designated time points for *mre11A470T* and *mre11A470T rap1-5* strains at 23˚C, 30˚C, and 37˚C.

https://doi.org/10.1371/journal.pone.0183549.g008

incorporation of heterodimers. This possibility has precedent from positive and negative effects on transcription elicited by histone modification within a microenvironment (e.g., [36, 37]).

### The heterodimer paradox

The lack of a phenotype generated from the ectopic locus is paradoxical since both genomic and ectopic alleles are expressed. *MRE11* at an ectopic site can complement genomic *mre11Δ* alleles, the combination of the genomic and ectopic loci increases the steady state level of transcription, and a Tap-tagged *MRE11* or *mre11A470T* can be expressed at the ectopic site. Proteins encoded by *MRE11* or *mre11A470T* at the genomic and ectopic sites are also equally capable of translation [S5 Fig]. For example, heterodimers must be different, depending upon the genomic allele, in order to produce only two possible phenotypes from all allelic

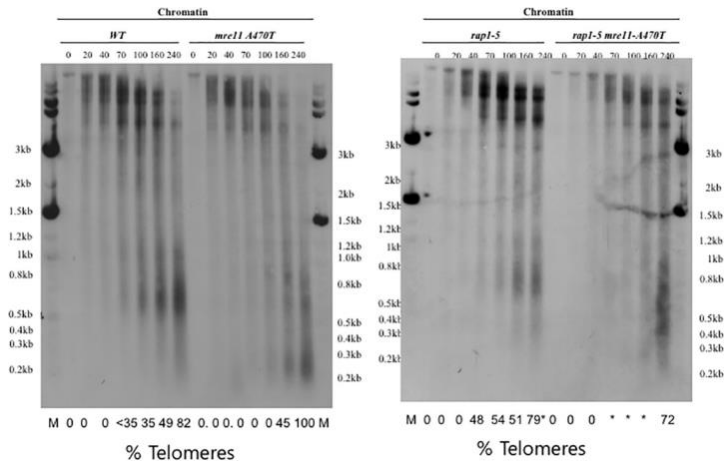

**Fig 9. Telosome behavior in *rap1-5 mre11A470T* cells.** Cells for telosome analysis were grown at 30˚C. Chromatin was isolated from *MRE1l*, *mre11A470T*, *rap1-5*, and *mre11A470T rap1-5* and digested with one unit /100 μl MNase for the indicated time [in seconds] by Southern analysis using telomere sequences as the probe. Both *mre11A470T* and the double mutant behave similarly at 30˚C. We cannot rule out some subtle difference between these two strains. An asterisk refers to values greater than 0 but having insufficient pixels for quantification.

https://doi.org/10.1371/journal.pone.0183549.g009

combinations. Specifically, the e-*mre11A470T* g-*MRE11* (1) and e-*MRE11* g-*mre11A470T* (2) should both produce a mre*11A470T/ MRE11* heterodimer. If so, they have distinct phenotypes. Possibly, Mre11/Rad50 interaction, 3' end formation, or heterodimer instability interfere with the production of heterodimers.

## Conservative segregation, the heterodimer problem and heritability

A conservative segregation pattern is a general requirement for telomeric chromatin, regardless of the specific mechanisms involved. There is precedence for conservative segregation or spreading of heterochromatin that is caused by specific base modifications [57].

In order to explain our results, we must assume that heterodimers are dysfunctional in an activity or in stability. Alternatively, there may be a strong selection for homodimer formation on telosomes. Telomeres have non-redundant mechanism of telomere cohesion, and may be specific to sister chromatid cohesion. Thus, separation occurs first at G2. Possibly, alignment of sister chromatids contributes for cohesin-mediated transient sister chromatid separation in G2 [Fig 10]. Alternatively, a closer proximity of the sister alleles may help to maintain the pre-existing state.

## Characterization of *Mre11A470T*

The *mre11A470T* allele confers a unique chromatin structure. The telosome that contains the Mre11A470T protein requires longer MNase incubation for the release of the mutant particle from the total chromatin. Also, relative to wild type, the distribution is more diffuse than expected. This distribution suggests conformational changes that result in promiscuous subtelomeric MNase cut sites. However, this argument oversimplifies the mechanism since we have identified an expected wild type phenotype that is diffuse. The source of this property remains unknown. However, one speculative mechanism is a region of lowered stability or altered conformation, such as a fold-back structure [38], formed between an *mre11A470T* telomere and the subtelomeric region. The fold-back structure may have a lowered fidelity of MNase cut sites, and as a consequence, a broad range of telomere sizes after digestion.

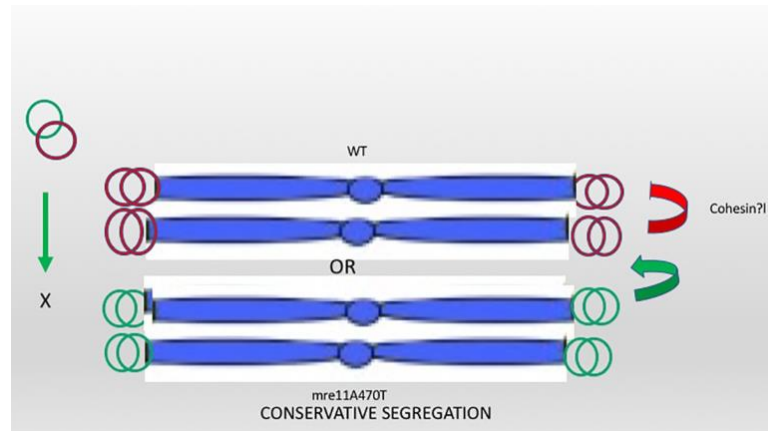

**Fig 10. Either Mre11 or Mre11A470R telosomes sequester to replicated sister chromatids.** This model shows the presence of only one of two possible configurations after replication depending on the genomic locus. Each pair of sister chromatids have identical telosomes [red or green] at both telomeres. Exchange between chromatids is shown by the green arrow. A sister chromatid specific mechanism of telomere separation in G2 may be mediated by a unique mechanism that is essential for subsequent separation [58]. The choice of ectopic vs genomic allele may be the consequence of an irreversible dimerization process that is established before integration of the ectopic allele.

https://doi.org/10.1371/journal.pone.0183549.g010

### Rad50/Mre11 interactions

One of the reasons for our interest in the Mre11A470T protein is its proximity to Mre11/Rad50 interface in eukaryotic crystal structures [39]. Indeed, the "dominant" epistatic pattern is a common feature of the A470 alleles. In each case, a wild-type gene integrated at an ectopic site failed to produce wild-type telomeres. This feature suggests their common involvement of the A470 motif in related functions, possibility regulating the Mre11A470T/Rad50 structure.

### Synergistic effects of *rap1-5* and *mre11A470T*

The *rap1-5 mre11A470T* double mutant is synergistic at all temperatures in growth rate, and at 37°C in viability. Rap1 and Mre11 appear to act on the same substrate through differing pathways, both of which affect growth rate. Pathways are less clear in the telosome profiles at 30°C, in which *mre11A470T* is epistatic to *rap1-5*. Mre11 may play a role in the equilibrium between Tel1 and Rif1 and in size homeostasis [16, 59]. Mre11 may participate in the formation of homeostasis, while Rap1 may play a role in protection against non-homologous end-joining [60, 61]. Experiments that examine the dynamics of wild-type strains are needed to resolve this issue.

### Supporting information

**S1 Fig. To investigate effect of *RAD5* on telomere size, *RAD5* was deleted in *MRE11* and *mre11A470T* strains.** *Xho*I digestion of DNA from the following genotypes were tested: Lane1, *MRE11 RAD5*; lane2, *mre11A470T RAD5*; lane3, *MRE11rad5*Δ; and lane 4 is *mre11A470T rad5*Δ strains. The *rad5*Δ deletion did not give rise to a telomere size change. The *mre11A470T rad5Δ* double mutant has the same short telomere phenotype and heritability as *mre11A470T* cells.
(TIF)

**S2 Fig. Telosome derived telomere size.** We extracted DNA from WT, *mre11A470T*, *mre11Δ*, *mre11A470T rap1-5*, and *mre11Δ rap1-5* chromatin. The DNA was digested with *Xho*I and subjected to Southern blot analysis using the A750 probe. Two exposures are shown to visualize all fragments.
(TIF)

**S3 Fig. Telomere sequencing in *mre11A470T* strains.** We sequenced multiple X- class telomeres from independent clones of wild-type or *mre11A470T* cells using the C-tail method [51] with the X-class primers adjacent to a telomere tract. The X-class telomeres represent a group of 13 telomeres that have identical junctions with the telomere. S3A: Sequence of three clones of a *mre11A470T* was conducted on independent telomeres derived from a single source. The color code is present simply for convenience and is not meant to indicate causality.
(TIF)

**S4 Fig. Multiple clones shown at telomere-proximal and distal sequences of a wild-type strain.** The top sequences run from the telomere proximal to distal direction. The color code denotes the type of misalignment or gap among different telomeres even when they were in the minority. The comparisons are not used to determine causal relationships, simply the type of events that are occurring.
(TIF)

**S5 Fig. Cycloheximide stability assays.** [A] Stability assay for Mre11-Tap and Mre11A470T-A portion of the CHX stability assay of strains containing a locus at the genomic site and TAP-tagged derivatives [Mre11-TAP and Mre11A470PT-TAP}. Westerns were probed with anti-TAP antibody antibodies as described in part B. The protein migrates more slowly due to the TAP tag. Size markers are provided on the left. The blot was stripped and re-probed with anti-acting antibody. [B] CHX protein stability assay was performed in *MRE11* and *mre11A470T* strains. Cells were grown in YPD to mid-log phase, and 100 μg of cycloheximide was added. Cells were incubated and collected at designated time point. The proteins were prepared, and Mre11 antibody was used to detect Mre11 and Mre11A470T protein. Mre11A470T has the same protein stability as Mre11. Actin was observed after stripping the blot and probing with an anti-actin antibody.
(TIF)

**S6 Fig. Mre11 RNA concentrations in genomic versus ectopic sites.** To test the transcriptional expression level of overall Mre11 in strains carrying two loci [genomic and ectopic], real time-qPCR of the reverse transcribed RNA was conducted as described in Materials and Methods. The first, second, and third bars refer to WT [*g-MRE11*] [as a control], *e-MRE11 g-MRE11* and *e- mre11A470T g-MRE11* strains, respectively. The expression of both e-*MRE11* g-*MTE11* and *e*-mre11A470T g-MRE11 were three- to four-fold greater than strains having only a single genomic locus. The Y axis represents the fold increase in transcripts relative to wild type.
(TIF)

## Acknowledgments

We would like to thank Drs. Alpana Kumari and Danielle Tatum for their early work in these studies [and for the data that led to S5 Fig] and Dr. Astrid Engel and Bonnie Hoffman for critical reading of the manuscript. We also thank Drs. Martin Kupiec, Samuel Landry, Edward Lewis, Thomas D. Petes, Raymund Wellinger, and Virginia Zakian for helpful comments during this project.

## Author Contributions

**Conceptualization:** Arthur J. Lustig.

**Data curation:** In-Joon Baek, Daniel S. Moss, Arthur J. Lustig.

**Formal analysis:** In-Joon Baek, Daniel S. Moss, Arthur J. Lustig.

**Funding acquisition:** Arthur J. Lustig.

**Investigation:** In-Joon Baek, Daniel S. Moss, Arthur J. Lustig.

**Methodology:** In-Joon Baek, Daniel S. Moss.

**Project administration:** Arthur J. Lustig.

**Resources:** Daniel S. Moss, Arthur J. Lustig.

**Software:** In-Joon Baek, Arthur J. Lustig.

**Supervision:** Daniel S. Moss, Arthur J. Lustig.

**Validation:** In-Joon Baek, Daniel S. Moss.

**Visualization:** In-Joon Baek.

**Writing – original draft:** Arthur J. Lustig.

**Writing – review & editing:** In-Joon Baek, Daniel S. Moss, Arthur J. Lustig.

## 1st qPCR analysis

| Background Cq | WT Cq | mre11A470T Cq | WT ΔCq | mre11A470TΔCq | WT ΔCq fold enrichment | mre11A470T ΔCq fold enrichment |
|---|---|---|---|---|---|---|
| 7.19 | 24.32 | 26.39 | 17.13 | 19.2 | 6.97198E-06 | 1.66044E-06 |
| 11.14 | 25.58 | 27.5 | 14.44 | 16.36 | 4.49911E-05 | 1.18891E-05 |
| | | | WT Mean ΔCq fold enrichment | mre11A470T Mean ΔCq fold enrichment | WT Mean ΔCq fold enrichment Srd.Dev. | mre11A470T ΔCq fold enrichment Srd.Dev. |
| | | | 2.59816E-05 | 6.77478E-06 | 2.68836E-05 | 7.23276E-06 |
| | | | WT ΔΔCq fold enrichment (normalize to Δmre11) | mre11A470T ΔΔCq fold enrichment (normalize to Δmre11) | WT ΔΔCq fold enrichment Srd.Dev. (normalize to Δmre11) | mre11A470T ΔΔCq fold enrichment Srd.Dev. (normalize to Δmre11) |
| | | | 10.09691548 | 2.632804 | 10.44746585 | 2.810784475 |

| Background Cq | | Δmre11 Cq | Δmre11 ΔCq | | Δmre11 ΔCq fold enrichment | Δmre11 Mean ΔCq fold enrichment | Δmre11 ΔCq fold enrichment Srd.Dev. |
|---|---|---|---|---|---|---|---|
| 11.14 | | 29.27 | 18.13 | | 3.48599E-06 | 2.57322E-06 | 1.29086E-06 |
| 7.19 | | 26.59 | 19.4 | | 1.66044E-06 | | |
| | | | Δmre11 ΔΔCq fold enrichment (normalize to Δmre11) | Δmre11 ΔΔfold enrichment Stad.Dev.(normalize to Δmre11) | | | |
| | | | 1 | 0.501650913 | | | |

### Relative Activity

| | | |
|---|---|---|
| WT | 10.09691548 | |
| mre11A470T | 2.632804 | |
| Δmre11 | 1 | |

**1st qPCR raw graph**

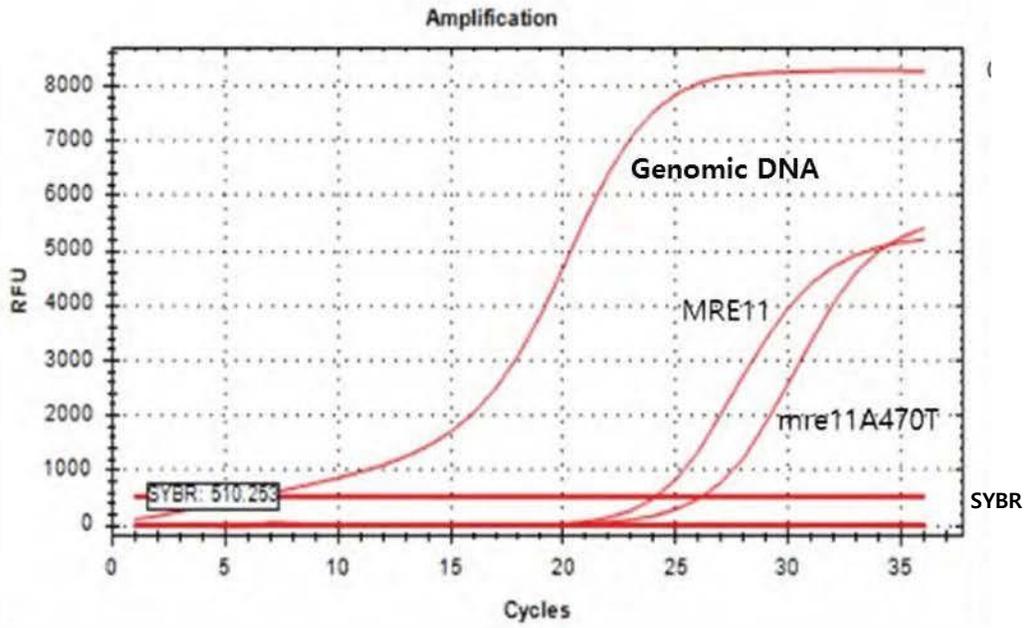

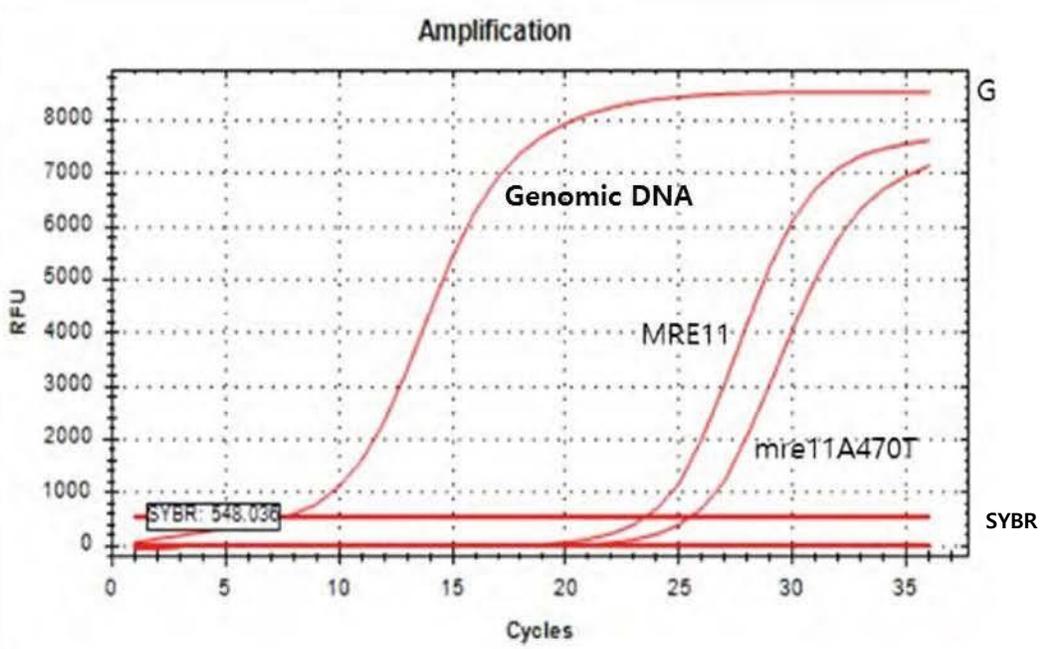

## 2nd qPCR analysis

| Background | WT Cq | mre11A470T Cq | WT ΔCq | mre11A470T ΔCq | WT ΔCq fold enrichment | mre11A470T ΔCq fold enrichment |
|---|---|---|---|---|---|---|
| 10.24 | 19.16 | 20.3 | 8.92 | 10.06 | 0.002064488 | 0.000644291 |
| 10.23 | 19.2 | 20.33 | 8.97 | 10.1 | 0.001994164 | 0.000911165 |
| | | | WT Mean ΔCq fold enrichment | mre11A470T Mean ΔCq fold enrichment | WT ΔCq fold enrichment Srd.Dev. | mre11A470T ΔCq fold enrichment Srd.Dev. |
| | | | 0.002029326 | 0.000777728 | 4.97266E-05 | 0.000188708 |
| | | | WT ΔΔCq fold enrichment (normalize to Δmre11) | mre11A470T ΔΔCq fold enrichment (normalize to Δmre11) | WT ΔΔCq fold enrichment Srd.Dev. (normalize to Δmre11) | mre11A470T ΔΔCq fold enrichment Srd.Dev. (normalize to Δmre11) |
| | | | 11.28726947 | 4.325783082 | 0.276583261 | 1.049610979 |

| Background | Δmre11 Cq | | Δmre11 ΔCq | Δmre11 ΔCq fold enrichment | Δmre11Mean ΔCq fold enrichment | Δmre11 ΔCq fold enrichment Srd.Dev. |
|---|---|---|---|---|---|---|
| 10.24 | 25.13 | | 14.89 | 3.29354E-05 | 0.000179789 | 0.000207682 |
| 10.23 | 21.81 | | 11.58 | 0.000326642 | | |
| | | | Δmre11 ΔΔCq fold enrichment (normalize to Δmre11) | Δmre11 ΔΔCq fold enrichment Stad.Dev. (normalize to Δmre11) | | |
| | | | 1 | 1.155144613 | | |

| | Relative Binding Activity |
|---|---|
| WT | 11.28726947 |
| mre11A470T | 4.325783082 |
| Δmre11 | 1 |

2nd qPCR raw graph

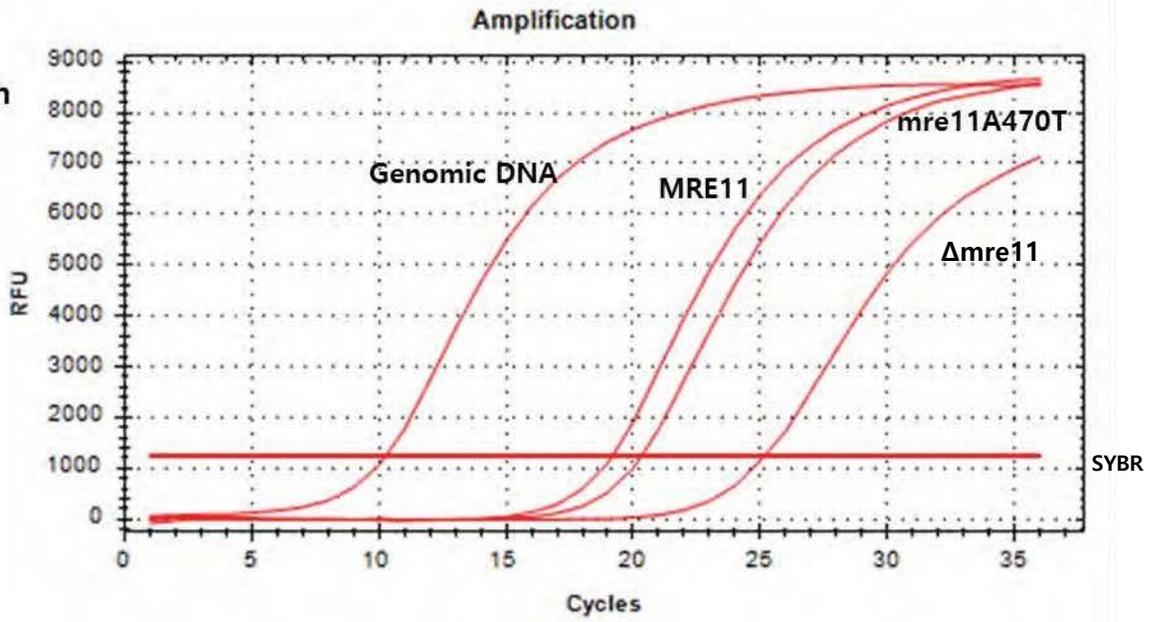

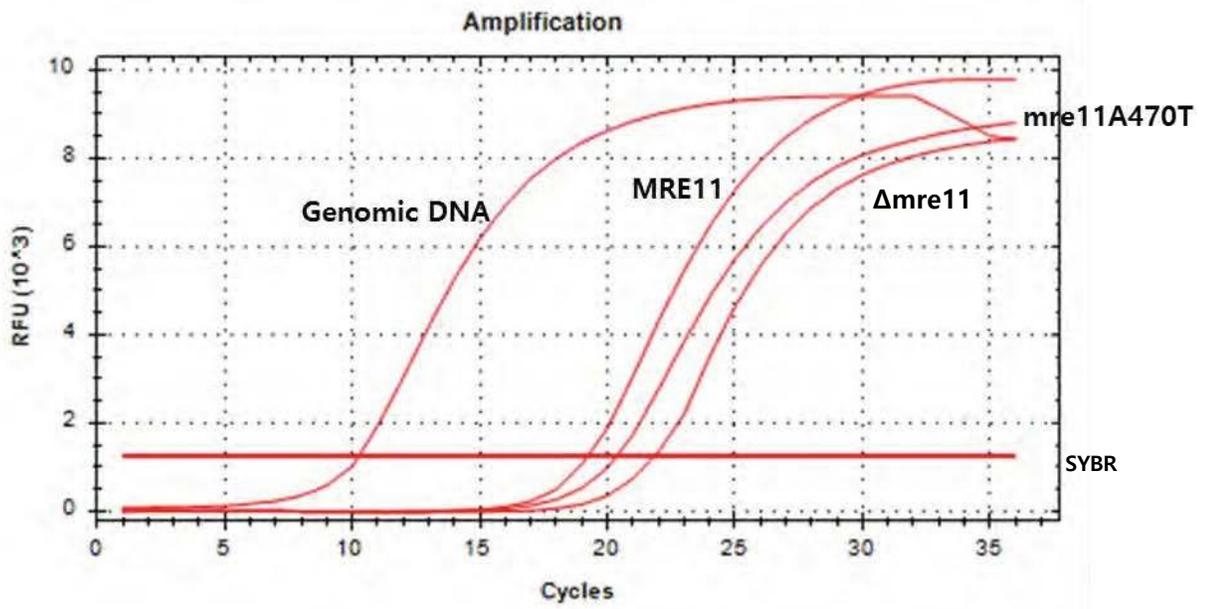

| Background | WT Cq | mre11A470T Cq | WT ΔCq | mre11A470T ΔCq | WT ΔCq fold enrichment | mre11A470T ΔCq fold enrichment |
|---|---|---|---|---|---|---|
| 10.21 | 24.32 | 26.39 | 14.11 | 16.18 | 5.65544E-05 | 1.3469E-05 |
| 10.42 | 25.58 | 27.5 | 15.16 | 17.08 | 2.7314E-05 | 7.21785E-06 |
| | | | WT Mean ΔCq fold enrichment | mre11A470T Mean ΔCq fold enrichment | WT ΔCq fold enrichment Srd.Dev. | mre11A470T ΔCq fold enrichment Srd.Dev. |
| | | | 4.19342E-05 | 1.03434E-05 | 2.06761E-05 | 4.42022E-06 |
| | | | WT ΔΔCq fold enrichment (normalize to Δmre11) | mre11A470T ΔΔCq fold enrichment (normalize to Δmre11) | WT ΔΔCq fold enrichment Srd.Dev. (normalize to Δmre11) | mre11A470T ΔΔCq fold enrichment Srd.Dev. (normalize to Δmre11) |
| | | | 19.90673877 | 4.910157092 | 9.815232065 | 2.098336722 |

| Background | | | Δmre11 Cq | Δmre11 ΔCq | Δmre11Mean ΔCq fold enrichment | Δmre11 ΔCq fold enrichment Srd.Dev. |
|---|---|---|---|---|---|---|
| 10.21 | | | 29.35 | 19.14 | 1.73096E-06 | 5.31148E-07 |
| 10.42 | | | 29.04 | 18.62 | 2.48211E-06 | 2.10653E-06 |
| | | | Δmre11 ΔΔCq fold enrichment (normalize to Δmre11) | Δmre11 ΔΔCq fold enrichment Stad.Dev. (normalize to Δmre11) | | |
| | | | 1 | 0.252143254 | | |

| | Relative Binding Activity |
|---|---|
| WT | 19.90673877 |
| mre11A470T | 4.910157092 |
| Δmre11 | 1 |

3rd qPCR raw graph

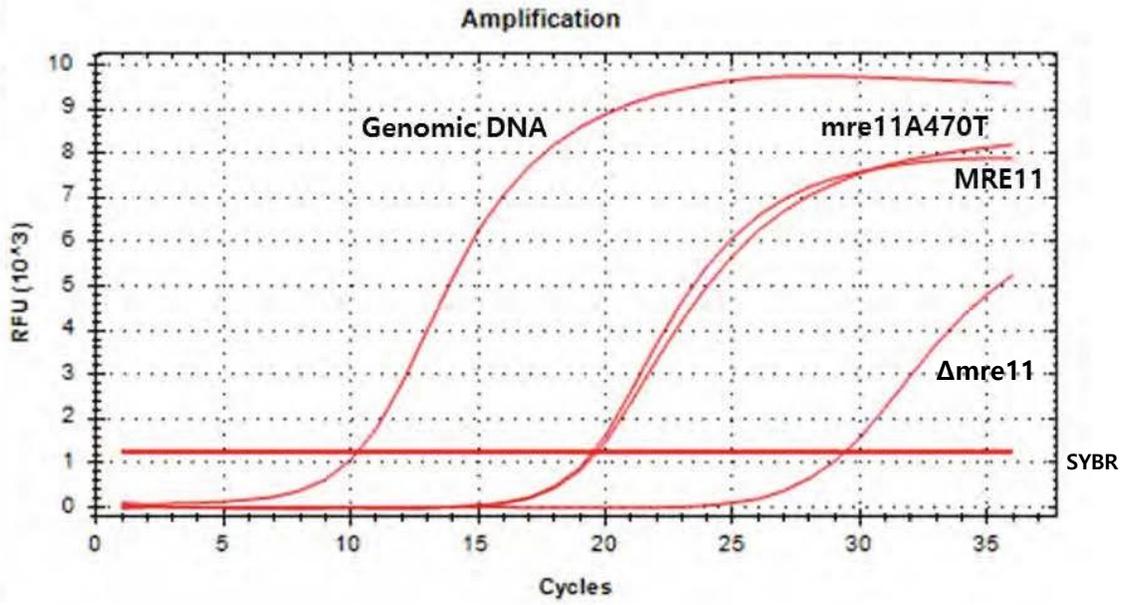

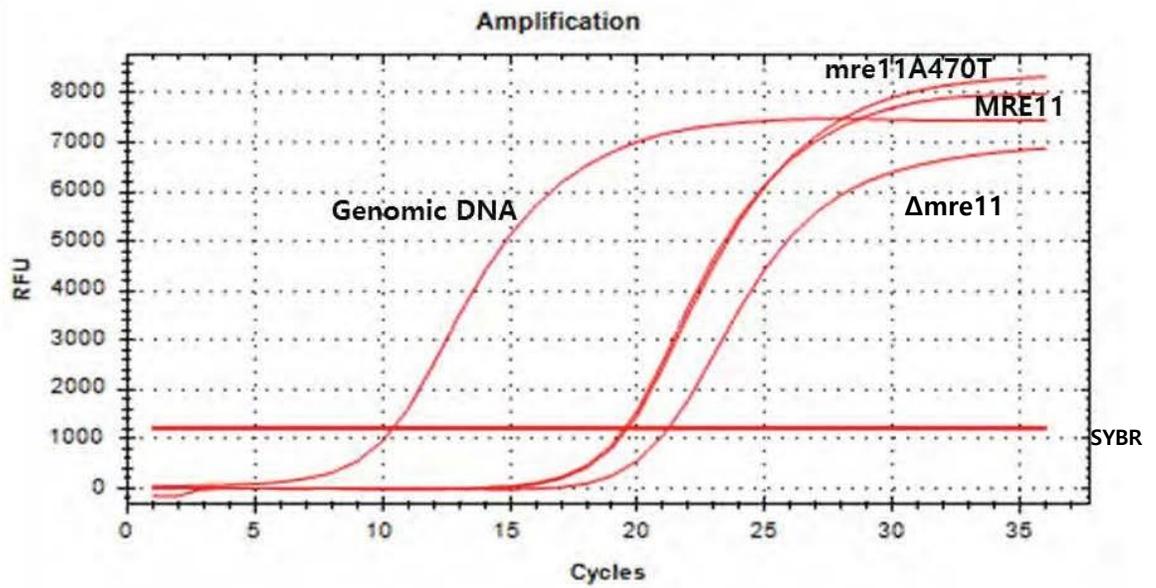

# EXPRESSION STUDIES

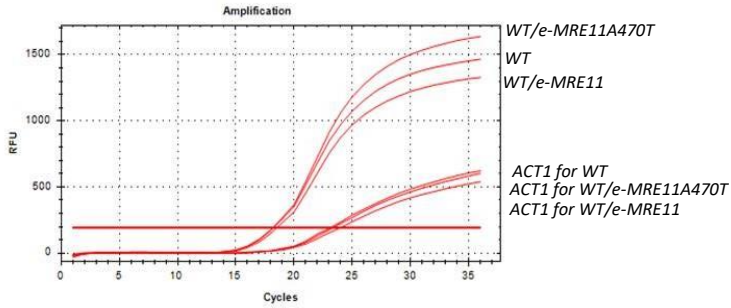

| Well | Fluor | Target | Content | Sample | Cq | |
|---|---|---|---|---|---|---|
| A01 | SYBR | | Unkn | | 18.49 | WT |
| A04 | SYBR | | Unkn | | 23.07 | ACT1 for WT |
| B01 | SYBR | | Unkn | | 18.94 | WT/e-MRE11 |
| B04 | SYBR | | Unkn | | 25.40 | ACT1 for WT/e-MRE11 |
| C01 | SYBR | | Unkn | | 17.64 | WT/e-MRE11A470T |
| C04 | SYBR | | Unkn | | 24.33 | ACT1 for WT/e-MRE11A470T |

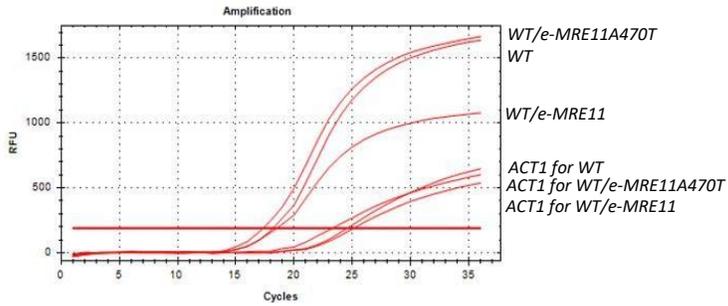

| Well | Fluor | Target | Content | Sample | Cq | |
|---|---|---|---|---|---|---|
| A02 | SYBR | | Unkn | | 18.18 | WT |
| A05 | SYBR | | Unkn | | 23.28 | ACT1 for WT |
| B02 | SYBR | | Unkn | | 18.48 | WT/e-MRE11 |
| B05 | SYBR | | Unkn | | 25.11 | ACT1 for WT/e-MRE11 |
| C02 | SYBR | | Unkn | | 17.32 | WT/e-MRE11A470T |
| C05 | SYBR | | Unkn | | 24.53 | ACT1 for WT/e-MRE11A470T |

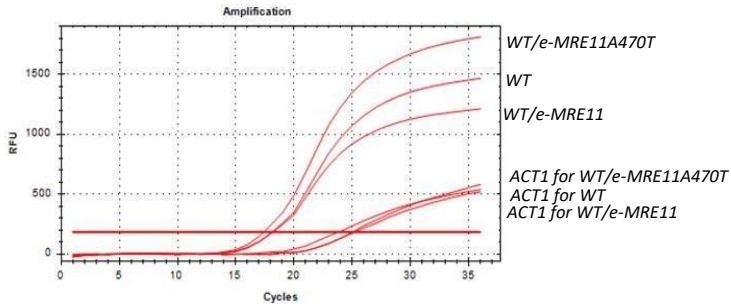

| Well | Fluor | Target | Content | Sample | Cq | |
|---|---|---|---|---|---|---|
| A03 | SYBR | | Unkn | | 18.17 | WT |
| A06 | SYBR | | Unkn | | 23.92 | ACT1 for WT |
| B03 | SYBR | | Unkn | | 18.21 | WT/e-MRE11 |
| B06 | SYBR | | Unkn | | 25.26 | ACT1 for WT/e-MRE11 |
| C03 | SYBR | | Unkn | | 17.44 | WT/e-MRE11A470T |
| C06 | SYBR | | Unkn | | 25.08 | ACT1 for WT/e-MRE11A470T |

Using the PCIA method, RNA was extracted from *WT, WT/ectopic MRE11,* and *WT/ectopic mre11A470T* strains.
The cDNA was synthesized from RNA.
Briefly, 4ul of iScript RT supermix( Bio-Rad) was added into 1ug of RNA and then fill dH2O up to 20ul.The complete reaction mix incubated
priming : 5min at 25 degree
Reverse transcription : 30min at 42 degree
RT inactivation : 1min at 95 degree
The tube placed on ice for 3 min and 10ul dH2O was added.
For the qPCR, iTaq™ Universal SYBR® Green Supermix (Bio-Rad) and CFX96 Touch™ Real-Time PCR Detection System (Bio-Rad) was used.

## ΔΔCq calculation method

| ACT1 Cq | WT Cq | WT ΔCq | WT ΔCq expression | WT ΔCq expression Std.Dve. | WT ΔΔCq expression | WT ΔΔCq expression Std.Dve. |
|---|---|---|---|---|---|---|
| 23.07 | 18.49 | -4.58 | 23.91758798 | 12.39524379 | 1 | |
| 23.28 | 18.18 | -5.1 | 34.2967508 | | 1 | |
| 23.92 | 18.17 | -5.75 | 53.81737058 | | 1 | |

| ACT1 Cq | E-MRE11 Cq | E-MRE11 ΔCq | E-MRE11 ΔCq expression | E-MRE11 ΔCq expression Std.Dve. | E-MRE11 ΔΔCq expression | E-MRE11 ΔΔCq expression Std.Dve. |
|---|---|---|---|---|---|---|
| 25.4 | 18.94 | -6.46 | 88.03467636 | 18.9145305 | 3.680750602 | 0.504913261 |
| 25.11 | 18.48 | -6.63 | 99.04415959 | | 2.887858391 | |
| 25.26 | 18.21 | -7.05 | 132.5139103 | | 2.462288827 | |

| ACT1 Cq | E-mre11A470T Cq | E-mre11A470T ΔCq | E-mre11A470T ΔCq expression | E-mre11A470T ΔCq expression Std.Dve. | E-mre11A470T ΔΔCq expression | E-mre11A470T ΔΔCq expression Std.Dve. |
|---|---|---|---|---|---|---|
| 24.33 | 17.64 | -6.69 | 103.2501452 | 39.31084297 | 4.316912946 | 0.287821073 |
| 24.53 | 17.32 | -7.21 | 148.0560875 | | 4.316912946 | |
| 25.08 | 17.44 | -7.64 | 199.4661324 | | 3.706352248 | |

Cq values were comes from CFX Manager software.
Base on the Cq values, ΔCq(Target Cq- reference Cq), ΔCq expression($2^{-\Delta Cq}$), mean ΔCq expression(average replicates), ΔCq expression Std Dev(standard deviation replicates), and ΔΔCq (normalize to Δmre11 for Real-Time qPCR, normalize to WT for RT-qPCR) was calculated.

**Step 1.** Normalize to (REF) : ΔCq(Target Cq- reference Cq)
**Step 2.** Exponential expression transform : ΔCq expression($2^{-\Delta Cq}$)
**Step 3.** Average replicates and calculate standard deviation :
mean ΔCq expression(average replicates), ΔCq expression Std Dev(standard deviation replicates)
**Step 4.** Normalize to Control : ΔΔCq

|  | Relative Expression |  | Relative Expression |
| --- | --- | --- | --- |
| WT mean | 1 | WT Stan Dev | 0 |
| E-MRE11 mean | 3.595524867 | E-MRE11 Stan Dev | 0.504913261 |
| E-mre11A470T mean | 4.023614097 | E-mre11A470T Stan Dev | 0.287821073 |
|  |  |  |  |
| WT Stan Err | 0 |  |  |
| E-MRE11 Stan Err | 0.291511807 |  |  |
| E-mre11A470T Stan Err | 0.166173574 |  |  |

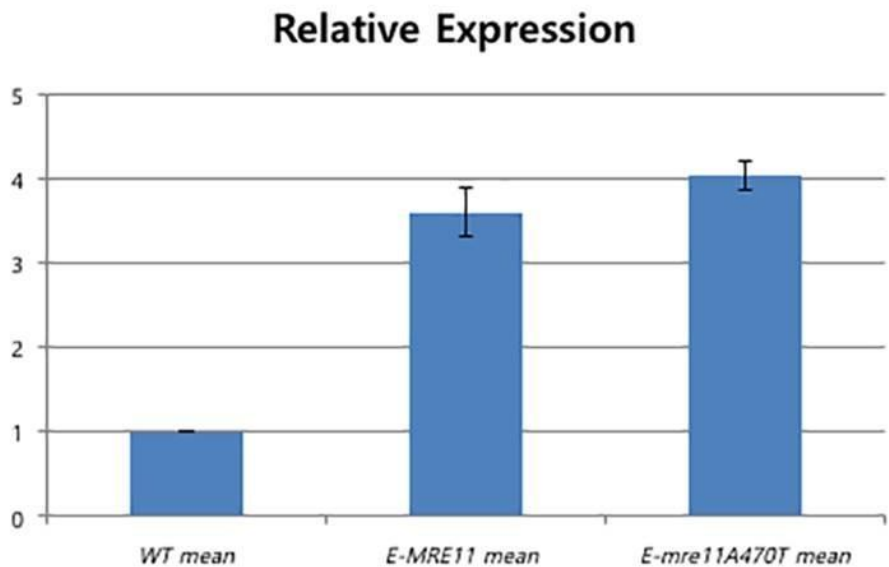

Each mean value, standard deviation, and standard error were calculated for bar graph.